\documentclass{article}

\usepackage{arxiv}

\usepackage[utf8]{inputenc} 
\usepackage[T1]{fontenc}    
\usepackage{hyperref}       
\usepackage{url}            
\usepackage{booktabs}       
\usepackage{amsfonts}       
\usepackage{nicefrac}       
\usepackage{microtype}      
\usepackage{lipsum}
\usepackage{graphicx}
\usepackage{multirow}
\usepackage{epstopdf}
\usepackage{subfigure}
\usepackage{amsmath}

\title{Design and experimental study of Tesla's thermomagnetic engine}

\author{
  Alex Estupiñán\thanks{djlexes@gmail.com, personal email address}\\
  Facultad de Ciencias, Escuela de Física\\
  Universidad Industrial de Santander (UIS)\\
  Colombia, Santander, Bucaramanga \\
  \texttt{alex.estupinan@saber.uis.edu.co} \\
   \And
      Rubén Rodríguez \\
      Ingeniería Mecatrónica \\
     Universidad Autónoma de Bucaramanga (UNAB)\\
      Colombia, Santander, Bucaramanga\\
      \texttt{Rrodriguez107@unab.edu.co} \\
   \And
   Jhon Amaya \\
   Ingeniería Mecatrónica     \\
   Universidad Autónoma de Bucaramanga (UNAB)\\
   Colombia, Santander, Bucaramanga\\
   \texttt{Jamaya419@unab.edu.co} \\
}

\begin{document}
\maketitle

\begin{abstract}

The scientists have shown great interest in the search for alternative means to generate energy, which are not contaminants and generate significant damage to the environment. One of the quite viable possibilities for this is to consider the construction of thermomagnetic motors, using mainly ferromagnetic materials. These materials are those that for a given temperature value; these lose the magnetic properties they have, that is, to be paramagnetic they become completely diamagnetic during a certain period of exposure to heat.

With the objective of demonstrate the Curie's law applied to this type of materials, we designed the model of an engine that works based on this law, to achieve this the tests of the running system were taken, which were filmed and then analyzed using the program Tracker Video Analysis and Modeling Tool for Physics Education.

In this paper, we present the results related to the magnetic and thermodynamic study of the efficiency of a motor designed by us, with the aim of showing the validation of Curie's law for iron and also being able to obtain the representative characteristics of this material such as magnetization and the Curie's constant using an experimental method.
	
\end{abstract}


\section{Introduction}

The experiment that was elaborate in this work, is based mainly on the efficiency in the properties of certain materials, which have a large amount of “free electrons”, which makes them susceptible to changes in temperature, electrical impulses, magnetic fields, among others, for example to be subjected to some heat interaction is intended to show the changes in the magnetization of these materials.

Our article explains the operation of the Curie's principle, which is based on the existence of a certain thermal point in which paramagnetic materials become diamagnetic, this temperature value was discovered by the french physicist Pierre Curie in 1895 \cite{revision}. With the above, the operation can be demonstrated in a practical environment, by heating a section of the paramagnetic material, which will be exposed to an electromagnetic field, where this material located on a rotating disk will lose its magnetic properties in the section exposed to heat becoming diamagnetic and losing the influence of the magnetic field only in that part. This will become visible with a disk rotation.

The remaining part of this paper is organized as follows: Section \ref{model}, describes the theoretical model corresponding to calculate the magnetization and the Curie constant of the material. In Section \ref{experimental}, shown the experimental procedure for the data-taking of experiment. In Section \ref{results},  presents the results obtained using the theoretical and experimental physical model presented in this artcile.

\section{Theoretical Background}
\label{model}

\begin{figure}[h!]
	\centering
	\subfigure[]
	{
		\includegraphics[width=0.47\linewidth]{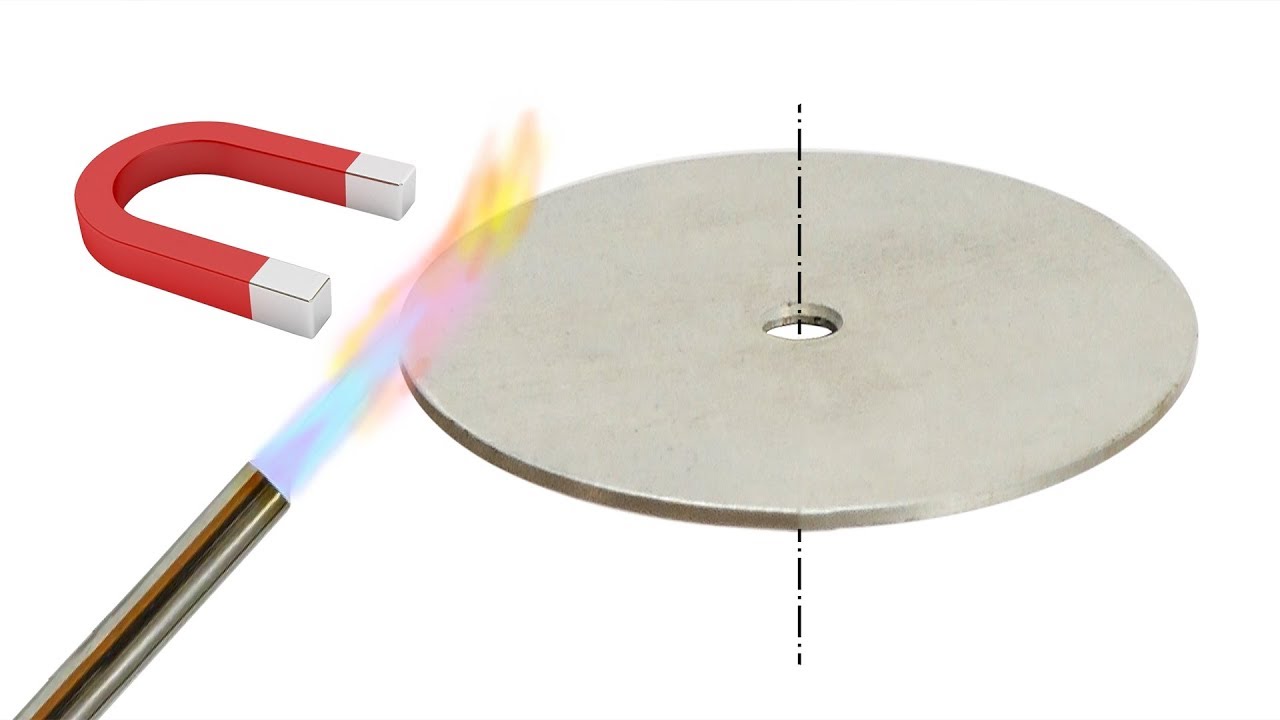}
	}
	\subfigure[]
	{
		\includegraphics[width=0.49\linewidth]{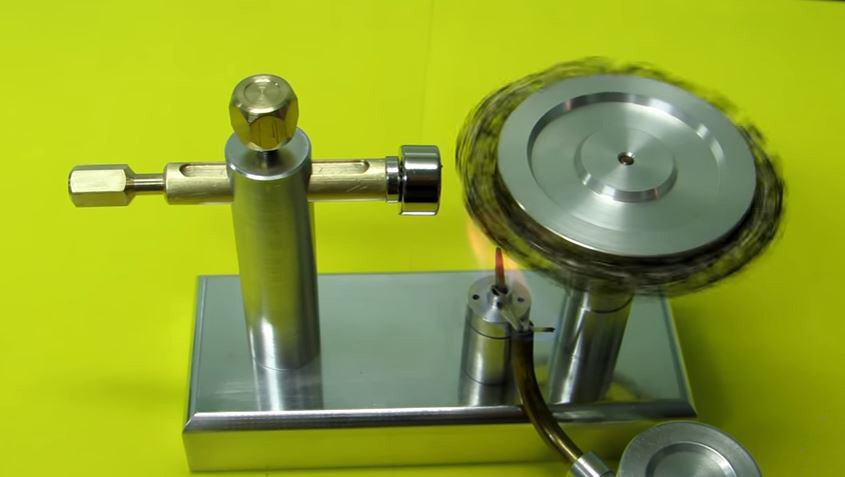}
	}
	\caption{(a) Main components of the physical system. (b) Experimental setup of the thermomagnetic engine.}
		\label{fig_1a}
\end{figure}

The physical system to be studied is shown in the Figure \ref{fig_1a}.

To begin to study the physical phenomenon presented in this work, we must find the most relevant factors of this, one of the main parameters to measure in this experiment should be the magnitude of the magnetic field generated by the toroidal coil, where its calculation is expressed in the Equation (\ref{eq1}).

\begin{equation}
B = \frac{\mu N i}{2 \pi R},
	\label{eq1}
\end{equation}

where $\mu$ is the magnetic permeability of the material, $N$ is the number of turns of coil, $i$ is the current and $R$ is the radius of disk. Knowing that the current can be obtained, through the Larmor precession (See the Equation (\ref{eq1})) for a quantity $k$ of electrons, we have that \cite{paper_1}:

\begin{equation}
i = ke \omega, \quad i = \frac{k e^2 B}{2 \pi m} 
	\label{eq2}
\end{equation}

Being, $m$ the mass of electron, $k$ the atomic number of material, $\gamma$ is the Lorentz factor, $e$ the charge of electron and $B$ the measured field that will affect the material. Then in this study, Curie's law (See the Equation (\ref{eq3})) was taken into account with the purpose of being able to find the magnetization value of the material for the Curie temperature of the iron (material studied in this work) \cite{Toftlund}.

\begin{equation}
M = \frac{CB}{T},
\label{eq3}
\end{equation}

where $M$ is the magnetization, $C$ is the specific constant of each material, $B$ is the induced magnetic field and $T$ the temperature in Kelvin. In addition, we can use a mathematical expression that relates, the Curie's constant, with the type of atomic organization that has the material to study \cite{kitel}, which is shown in the Equation (\ref{eq4}).

\begin{equation}
    C = \frac{m^{2}m_v N}{3K},
	\label{eq4}
\end{equation}

Where $C$ is the Curie's constant, $m$ the magnetic moment of an $Fe$ atom, $N$ is the number of atoms per unit volume, $K$ the Boltzmann constant and $m_v$ the permeability of free space.

Taking into account the rotation of the disk, we can apply the second law of rotational Newton to this system, where the following equation is obtained:

\begin{equation}
	\tau = \frac{1}{2} mra_t,
	\label{eq5}
\end{equation}

$\tau$ is defined as the resulting torque applied to the system, $m$ is the disk mass and $a_t$ the tangential acceleration.

To make a study of the energy of the system, which is consistent with the movement and the increase of the temperature that the system is winning, the kinetic and thermal energy of the system can be expressed as shown below \cite{barnes},

\begin{equation}
	E_c = \frac{1}{2}mv^{2}, \quad E_T=\frac{3}{2}kT.
	\label{eq6} 
\end{equation}

Where $E_c$ and $E_T$ is the kinetic and thermal energy respectively,  $v$ is the rotational speed and $m$ the disk mass.

\section{Experimental study}
\label{experimental}

The methodology that was followed to perform the experiment's arrangement, and with this to be able to carry out the experimental assembly, in addition to the materials necessary to be able to have this experiment complete (See Figure \ref{fig_2a}), were:

\begin{figure}[h!]
	\centering
	\includegraphics[width=0.70\textwidth]{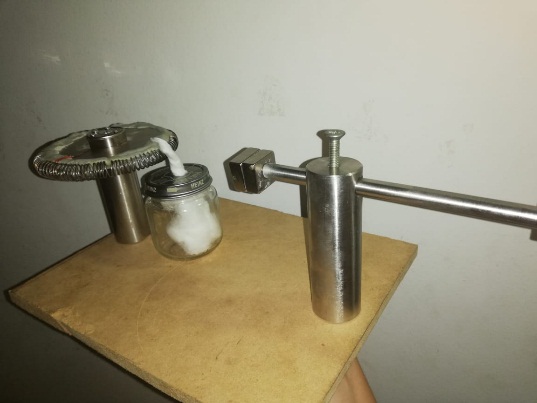}
	\caption{Experimental montage of the phenomenon to study.}
	\label{fig_2a}
\end{figure}

\begin{itemize}
	
	\item Design of the experimental model using the Solidworks program.
	
	\item 1 Hard disk bearing, due to its low friction.
	
	\item 0.4 meters Austenitic Stainless Steel 304.
	
	\item 5 meters of ferronickel wire.
	
	\item 1 alcohol-based lighter.
	
	\item Dynamic analysis using the Tracker program.

\end{itemize}

\section{Results}
\label{results}

So that we can begin to obtain results, related to the most important characteristics of the thermomagnetic motor, we will start by obtaining the value of the Curie's constant for the $ Fe $, using the Equation (\ref{eq4}), we have: 

\begin{equation}
	C = \frac{(18.54 \times 10^{-24})^{-2}\cdot 4 \pi \times 10^{-7}\cdot 8.5 \times 10^{28}}{3 \cdot (1.38 \times 10^{-23})} \nonumber
\end{equation}

\begin{equation}
	C = 0.89 \hspace{0.5em} \frac{A \cdot K}{m \cdot T}
	\label{eq7}
\end{equation}

Thus, assuming a metallic formation with momentum of $2$ Bohr magneton per atom and that the iron has a formation centered in the body (Body Centered Cubic structure, BCC) with a parameter of $0.286$ $nm$ \cite{kitel}, we obtained the result shown in the Equation (\ref{eq7}).

Continuing with the electrical study of the system, we can obtain the value of the current that passes through the toroidal laboratory, using the Equation (\ref{eq2}) and with a magnetic field of $25.7$ $mT$ and concentration $10^{7}$ electrons in the conductor:

\begin{equation}
	i = \frac{10^{7} \cdot (1.6 \times 10^{-19})^{2} \cdot 25.7 \times 10^{-2}}{2 \pi \cdot 9.1 \times 10^{-31}} \nonumber	 
\end{equation}

\begin{equation}
	i = 1.149 \hspace{0.5em} mA.
\end{equation}

In order to obtain the number of turns necessary for our experiment, we use the Equation (\ref{eq1}), as follows:

\begin{equation}
     N = \frac{2 \pi \cdot 3 \times 10 ^{-3} \cdot 25.7 \times 10^{-3}}{0.025 \cdot 1.149 \times 10^{-3}} \nonumber	
\end{equation}

\begin{equation}
	N = 168 \hspace{0.5em} turns.
\end{equation}

By experimentally measuring (using a teslameter) the value of the magnetic field of the coil as a function of distance, in addition using Equation (\ref{eq3}) to find the magnetization of the material, the data recorded in Table \ref{table_1a} were obtained.

\begin{table}[h!]
	\centering
	\begin{tabular}{cccllllll}
		\cline{1-3}
		\multicolumn{1}{|c|}{\textbf{Magnetic field [mT]}} & \multicolumn{1}{c|}{\textbf{Radial distance [mm]}} & \multicolumn{1}{c|}{\textbf{Magnetization  [$\mu$A/m]}} &  &  &  &  &  &  \\ \cline{1-3}
		\multicolumn{1}{|c|}{25.7} & \multicolumn{1}{c|}{1} & \multicolumn{1}{c|}{21.9} &  &  &  &  &  &  \\ \cline{1-3}
		\multicolumn{1}{|c|}{25.2} & \multicolumn{1}{c|}{3} & \multicolumn{1}{c|}{21.5} &  &  &  &  &  &  \\ \cline{1-3}
		\multicolumn{1}{|c|}{24.8} & \multicolumn{1}{c|}{5} & \multicolumn{1}{c|}{21.1} &  &  &  &  &  &  \\ \cline{1-3}
		\multicolumn{1}{|c|}{24.5} & \multicolumn{1}{c|}{7} & \multicolumn{1}{c|}{20.9} &  &  &  &  &  &  \\ \cline{1-3}
		\multicolumn{1}{|c|}{24.2} & \multicolumn{1}{c|}{8} & \multicolumn{1}{c|}{20.6} &  &  &  &  &  &  \\ \cline{1-3}
		\multicolumn{1}{|c|}{23.9} & \multicolumn{1}{c|}{9} & \multicolumn{1}{c|}{20.3} &  &  &  &  &  &  \\ \cline{1-3}
		\multicolumn{1}{l}{} & \multicolumn{1}{l}{} & \multicolumn{1}{l}{} &  &  &  &  &  & 
	\end{tabular}
	\caption{Experimental values for the magnetic field and the magnetization of Iron for different distances.}
	\label{table_1a}
\end{table}

\begin{figure}[h!]
	\centering
	\includegraphics[width=0.80\textwidth]{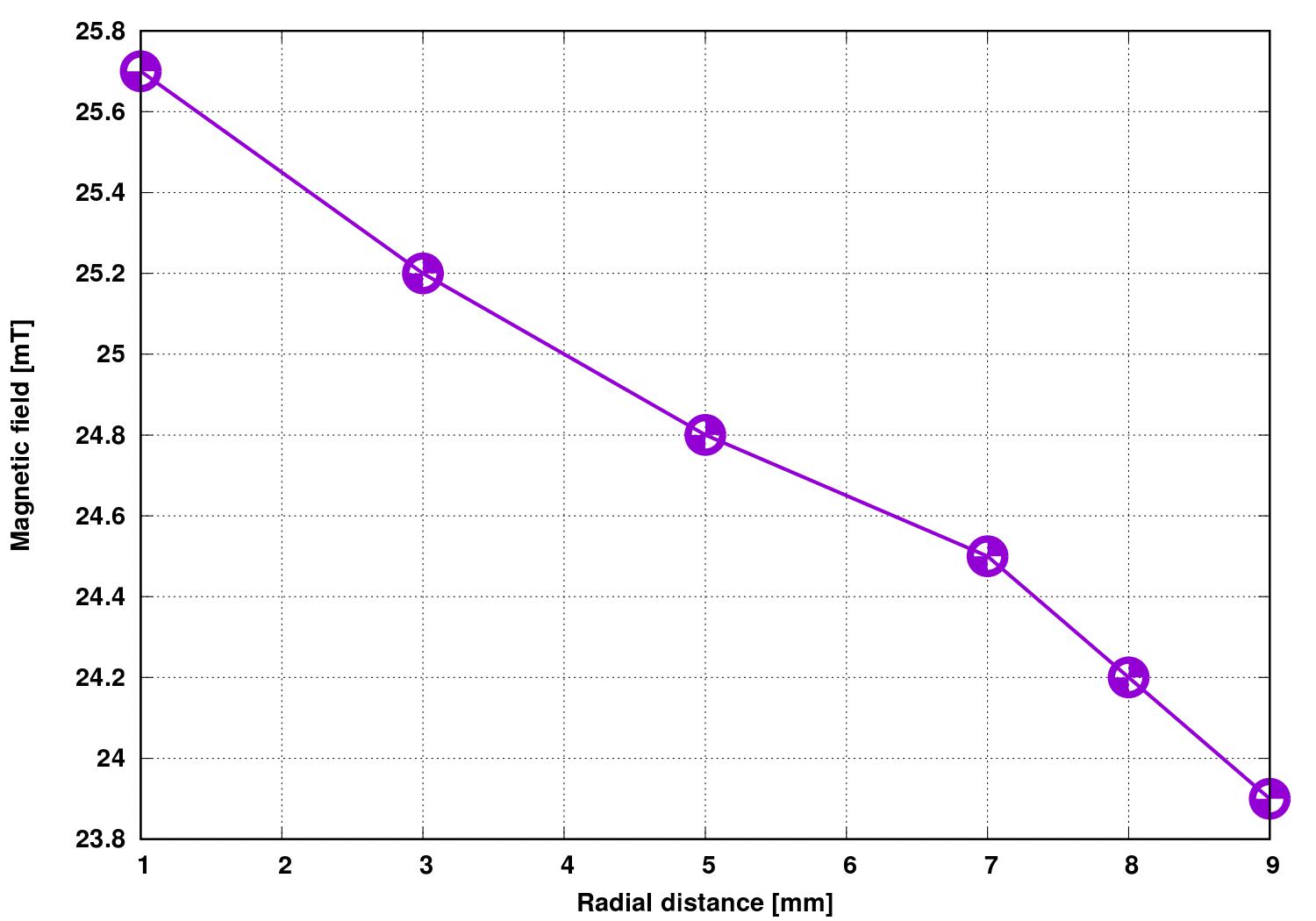}
	\caption{Experimental results for the magnetic field.}
	\label{campo}
\end{figure}

\begin{figure}[h!]
	\centering
	\includegraphics[width=0.80\textwidth]{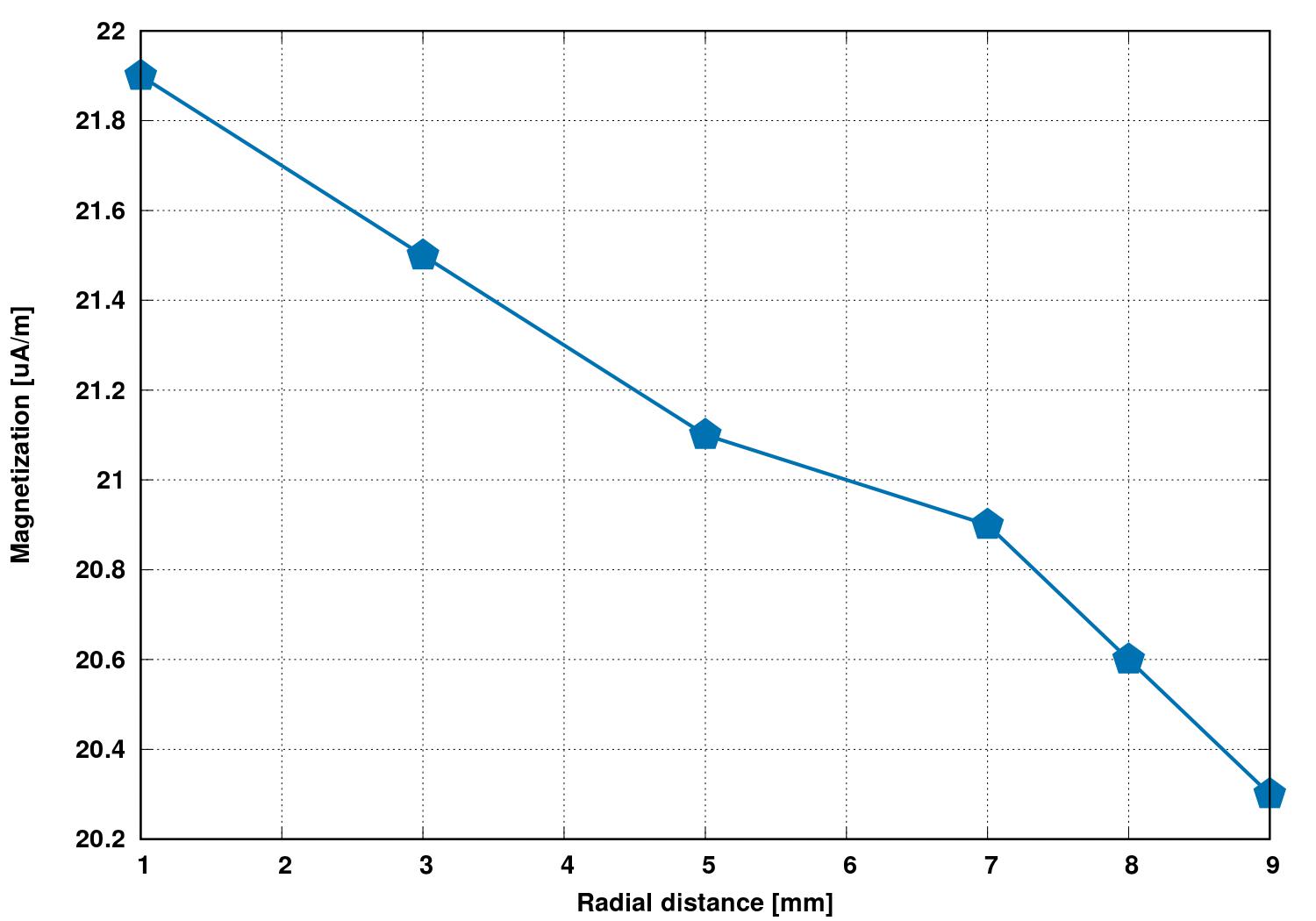}
	\caption{Experimental results for the magnetization.}
	\label{marnetization}
\end{figure}

Using the data shown in Table \ref{table_1a}, it was possible to obtain Figure \ref{campo} for the magnetic field and Figure \ref{marnetization} for the magnetization of the toroidal iron coil. In these graphs we can observe a dependence with the inverse of the distance, where it is also of great importance to note that the magnetization that we obtain for the Curie's temperature is of the order of $\mu A/m $, which is supported by the experiments in which the iron acquires diamagnetic characteristics, that is to say under these conditions the magnetization of the material will be almost.

\newpage

\begin{figure}[h!]
	\centering
	\includegraphics[width=0.80\textwidth]{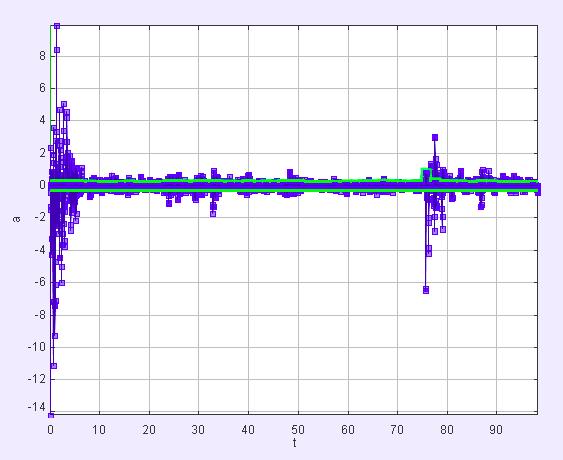}
	\caption{Tangential acceleration, the blue lines are the strong changes of speed in movement and the green lines indicate the average value.}
	\label{aceleracion}
\end{figure}

Continuing, with the obtaining of the results of the energy acquired by the system, taking into account that the mass is $30$ $g$, the radius of $10$ $cm$ and the color is blue of the flame, where its wavelength is $495$ $nm$ and average tangential acceleration of $1$ $cm/s^{2}$ (See the Figure \ref{aceleracion}), we give solution to equations (\ref{eq5}) and (\ref{eq6}), thus having:

\begin{equation}
	\tau = \frac{0.03 \cdot 0.1 \cdot 0.01}{2} \nonumber
\end{equation}

\begin{equation}
	\tau = 1.5 \times 10^{-5} \hspace{0.5em} J
\end{equation}

\begin{equation}
	E = \frac{h \cdot c}{\lambda},
\end{equation}

\begin{equation}
	E = \frac{6.63 \times 10^{-34} \cdot 3 \times 10^{8}}{495 \times 10^{-9}}, \nonumber
\end{equation}

\begin{equation}
	E = 4.018 \times -19 \hspace{0.5em} J, \quad E = 2.51 \hspace{0.5em} eV
\end{equation}

\begin{figure}[h!]
	\centering
	\includegraphics[width=0.80\textwidth]{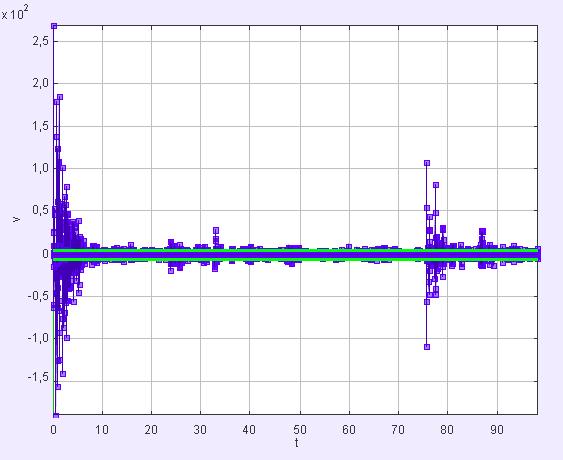}
	\caption{Tangential velocity, the blue lines are the strong changes of speed in movement and the green lines indicate the average value.}
	\label{velocity}
\end{figure}

On the other hand the kinetic and thermal energy, is the kinetic and thermal energy using the Equation (\ref{eq6}) and the Figure \ref{velocity}, in the following way:

\begin{equation}
	E_c = \frac{0.03 \cdot 0.2^{2}}{2},  \nonumber	
\end{equation}

\begin{equation}
	E_c = 0.6 \hspace{0.5em} mJ,
\end{equation}

where the average tangential velocity, measured using the Tracker analysis program, was $0.2$ $m/s$.

\begin{equation}
	E_T = \frac{3 \cdot 1.38 \times 10^{-23} \cdot 1043}{2}, \nonumber
\end{equation}

\begin{equation}
	E_T = 21.59 \times 10^{-21} \hspace{0.5em} J, \quad E_T = 134.9 \hspace{0.5em} meV.
\end{equation}

Where it should be noted that the Curie temperature for iron is $1043$ $K$.

Finally, we calculate the value of the efficiency of the engine studied in this work, this value is obtained as follows:

\begin{equation}
	E_{f} = 0.55	\cdot E_{f}(Fe), \quad E_{f} = 0.55	\cdot 0.49, \nonumber
\end{equation}

\begin{equation}
	 E_{f} = 0.27, \hspace{0.5em} \%E_{f} = 27 \%.
\end{equation}

With this final result, we can see that with our thermomagnetic motor we can obtain an efficiency of 27 \% in its operation.

\section{Conclusions}

As for the experimental part of the work, certain details must be taken into account, such as the source provided by the magnetic field, where in this case are the magnets, which must be at a very close distance coiled so that this have the greatest affectation of the possible surface thereof and in this way a more uniform movement can be produced. The bearing that holds the disc and the winding must have the least possible friction, so that it can rotate with more freedom, in this way, having the interaction of energy between the flame and the material is easier to interact, not have this freedom can cause a deformation of the material, if you do not have a uniformity of material throughout the disk, you will not have a constant speed because some parts will be more than others. Finally, it was possible to experimentally obtain the Curie point of the $Fe$ and its magnetization, which approaches zero,  in addition the value of the efficiency was obtained for the theromagnetic machine of 27 percent.

\section*{Acknowledgments}

The authors would like to thank the Universidad Autónoma de Bucaramanga (UNAB), for lend us the installations and materials for realizing this experiment presented in this paper.

\end{document}